\RequirePackage{fix-cm}
\documentclass[smallextended]{svjour3}       
\smartqed  
\usepackage{graphicx}
\usepackage{xcolor}
\usepackage{xfrac}
\usepackage{float}
\begin{document}

\title{\emph{Ariel} mission planning} 
\subtitle{Scheduling the survey of a thousand exoplanets}

\titlerunning{\emph{Ariel} mission planning}        

\author{J.C.~Morales$^{1,2}$ \and
        N.~Nakhjiri$^{1,2}$ \and
        J.~Colom\'e$^{1,2}$ \and
        I.~Ribas$^{1,2}$ \and
        E.~Garc\'ia$^{1,2}$ \and
        D.~Moreno$^{1,2}$ \and
        F.~Vilardell$^{1,2}$
}

\authorrunning{Morales et al.} 

\institute{ \at $^{1}$Institut de Ci\`encies de l'Espai (ICE, CSIC),
              Campus UAB, C/ de Can Magrans s/n, E-08193 Bellaterra, Spain
              \newline
              $^{2}$Institut d'Estudis Espacials de Catalunya (IEEC),
              C/ Gran Capit\`a 2-4, E-08034 Barcelona, Spain
              \newline
              \email{morales@ice.cat}
}

\date{Accepted: December 2021}

\maketitle

\begin{abstract}
Automatic scheduling techniques are becoming a crucial tool for the efficient planning of large astronomical surveys.
A specific scheduling method is being designed and developed for the Atmospheric Remote-sensing Infrared Exoplanet Large-survey (\emph{Ariel}) mission planning based on a hybrid meta-heuristic algorithm with global optimization capability to ensure obtaining satisfying results fulfilling all mission constraints. We used this method to simulate the \emph{Ariel} mission plan, to assess the feasibility of its scientific goals, and to study the outcome of different science scenarios.
We conclude that \emph{Ariel} will be able to fulfill the scientific objectives, i.e. characterizing $\sim$1000 exoplanet atmospheres, with a total exposure time representing about 75--80\% of the mission lifetime. We demonstrate that it is possible to include phase curve observations for a sample of targets or to increase the number of studied exoplanets within the mission lifetime. Finally, around 12--15\% of the time can still be used for non-time constrained observations.

\end{abstract}

\section{Introduction}
\label{intro}
Exoplanet transmission and emission spectroscopy has proven to be a unique tool to characterize exo-atmospheres. This is based on the determination of the transit and occultation depths as a function of wavelength, which yield the abundances of chemical species in the planet atmosphere, as well as its temperature profile (see \cite{Seager2010,Kaltenegger2017,Madhusudhan2019} for a review). The Atmospheric Remote-sensing Infrared Exoplanet Large-survey, \emph{Ariel}, is an European Space Agency (ESA) space mission under design study that will provide such kind of observations for a large number of exoplanets. Its main goal is the study of a statistical sample of exoplanet atmospheres to understand their structure and constituents \cite{Tinetti2018}. This might also provide information about the planet formation and evolution.

\emph{Ariel} consists of a 1-m class telescope whose main instrument is a low-resolution spectrograph. 
It is expected to be launched in 2028 and its nominal operations phase will last for 3.5\,years. During this period it will characterize the atmosphere of a sample of about 1000 exoplanets \cite{Edwards2019} with low-resolution spectrographs covering the infrared wavelengths bands from 1.1 to 7.8\,$\mu$m. To do so, it will follow-up two types of planetary events: transits, i.e. when the planet passes in front of the star; and occultations, i.e. when it passes behind. Observing such events are time-critical because they occur at specific times for each planetary system according to their ephemerides and their orbital properties. This imposes stringent constraints to the planning of the observations. Besides, \emph{Ariel} may also be useful to study the planet spectrum variability during its orbit around the host star (hereafter, phase curve) for several systems in order to study atmospheric circulation.

This objective represents a major challenge for the mission planning. Any planning should fulfill the science goals, i.e. characterize $\sim$1000 exoplanets, while optimizing the telescope time concurrently. Besides, mission constraints such as the visibility of the targets, the slew time, and the number of observations for each star need to be taken into account. The large number of targets, the time-criticality of planetary events, and the mission constraints make a manual computation of an optimized plan not affordable. 
The huge number of possible combinations makes automatic scheduling tools essential for the mission planning of surveys such as \emph{Ariel}.

A variety of different methods have been applied to build automatic tools for the scheduling of astronomical observations, from the more simple approaches like heuristic algorithms to sophisticated Artificial Intelligence (AI) techniques. Different methods have their own ups and downs. Heuristic methods can be described as algorithms that are based on a solving strategy. The strength of heuristic methods are based on their specific design. Some heuristics allow refinement of the plan after introducing new information, but some like greedy and hill-climbing algorithms optimize the plan once as they build it gradually with their best available option at the time. These local optimization techniques are usually fast but do a poor optimization in long-term scheduling. As an example, the Large Synoptic Survey Telescope scheduler is based on a greedy algorithm \cite{delgado2014lsst}. Heuristic methods can also be used to perform global optimization, like a neighborhood search approach \cite{yanez2003optimization}. On the other hand, a global optimizer is more reliable compared to a local one as it allows visiting of different local minimums in its process. Beside heuristics, AI provides a variety of global optimization methods. Evolutionary computation (EC, \cite{back1997handbook}), is a field in AI for global optimization, inspired by natural systems. The most commonly known algorithms of this field  are Genetic Algorithms (GAs) and Evolutionary Algorithms (EAs).

EAs are robust global optimization methods that work with the least assumption about the solution space. EAs search a larger area of the solution space than traditional heuristic methods, however, searching multiple locality and making no assumption about the solution space results in an increase of the computational cost of such methods. GAs are a more simple variation of EAs with a good optimization capability. An example of a GA used for scheduling of Hubble telescope is detailed in ref. \cite{sponsler1989genetic}. 
More advanced versions of EA are also used to solve the task of scheduling in the field of astronomy. For instance, this is the case of the Multi-Objective Evolutionary Algorithms (MOEA) used for the CARMENES survey \cite{GarciaPiquer2016,GarciaPiquer2017} and the James Webb Space Telescope \cite{giuliano2008multi}.

In the particular case of the \emph{Ariel} survey, we have developed a mission planning tool previously based on EA and GA \cite{GarciaPiquer2015} to a new tool that utilizes another sub category of EC algorithms, Swarm Intelligence \cite{beni1993swarm}, in addition to a specifically tailored meta-heuristic to schedule the mission, improving the computational cost. A detailed technical description and analysis of the algorithm will be presented in a forthcoming paper (Nakhjiri et al. in prep.).
Here, we focus on the application of this tool to simulate \emph{Ariel} mission plans fulfilling all constraints in order to study the feasibility of the science goals and different observational scenarios, and to analyze the characteristics of these plans.

In Sect.\,\ref{sec:ARIEL} we describe the sample of \emph{Ariel} targets we have used in this work, the operations that need to be scheduled, and the mission constraints that should be satisfied. In Sect.\,\ref{sec:algorithm} the algorithm used for the simulation is presented. In Sect.\,\ref{sec:results} we discuss the results of our mission planning simulations. Finally, conclusions are presented in Sect.\,\ref{sec:conclusions}.

\section{\emph{Ariel} scheduling constraints}
\label{sec:ARIEL}
A scheduling process is a constraint satisfaction problem that should plan one or different tasks fulfilling some restrictions and taking into account all possible limitations. Thus, any astronomical planning tool needs to be adapted to the kind of observations that should be planned, the scientific strategy, and the constraints. This includes, defining the targets that need to be observed, all the different operations performed and the constraints that must be satisfied. We provide further details in the following sections.

\subsection{Mission operations}
\label{subsec:operations}
As mentioned in the introduction, \emph{Ariel} will follow-up the transits and eclipses of exoplanet systems, and possibly the phase curve for some of them. These are the main tasks that need to be scheduled. However, other operations must be taken into account as well, for instance to maintain the spacecraft orbit or to perform auxiliary observations. These operations may have their own time constraints, which need to be defined, and they cannot overlap with each other. For the particular case of \emph{Ariel}, the mission operations that need to be considered by the scheduler are the following.

\paragraph{\textbf{Science observations.}}
They correspond to actual observations of exoplanet events. These are the most constraining tasks because they are time critical. Three different events are defined:
\begin{itemize}
    \item \underline{Transit}: when the planet passes in front of the star. 
    \item \underline{Occultation}: when the planet passes behind the star.
    \item \underline{Phase curve}: continuous follow-up of the planetary system during one or more orbital cycles. Its goal is to measure the planet spectral variability as a function of the orbital phase.
    \end{itemize}

Transit and occultation events are time constrained according to their ephemerides: the orbital period of the planet ($P$), a reference epoch and the duration from first to fourth contact ($T_{14}$), which are provided for each target (see Sect.\,\ref{subsec:sample}). As defined in the mission requirements, each of these events will be observed during 2.5$\times T_{14}$ around the central time of the transit or occultation to secure an accurate determination of the flux baseline before and after the planet crosses the star.

On the other hand, in order to facilitate the analysis of light curves and correct any possible systematic effect, in the current mission design, phase curves are also time constrained. Targets requesting such observations will be monitored from 0.75$\times T_{14}$ before the start of an occultation to 0.75$\times T_{14}$ after the end of another occultation including as many exoplanet orbital cycles as requested in between.

\paragraph{\textbf{Calibrations.}}
These tasks consist in the follow-up of stable bright G-type stars that are used to monitor and calibrate the response of the \emph{Ariel} instruments. They will be planned throughout the mission lifetime following a scheme designed to control the stability of the instrument at different timescales \cite{Pearson2020}. Therefore, although these operations are not time-critical, they should be planned in regular intervals. The current setup for \emph{Ariel} calibration operations allows for some scheduling flexibility:
\begin{itemize}
    \item \underline{Short calibrations}: 1\,hour every 36$\pm$12\,hours.
    \item \underline{Long calibrations}: 6\,hours every 30$\pm$10\,days.
\end{itemize}
Priority is given to long calibrations, i.e., short calibrations close to a long calibration are removed from the plan. On average, this calibration setup takes almost 300\,hours per year ($\sim$3\% of the mission lifetime). For the present exercise, a set of 536 stable G-type stars uniformly distributed over the sky are available, making sure that there is always a calibration source within $\sim$5\,deg of any exoplanet target.

\paragraph{\textbf{Station keeping operations.}} These operations are devoted to maintain the orbit of the spacecraft, and its cadence and duration will be decided by the spacecraft operators at ESA. As a preliminary approach, we assume that station keeping operations will take about 4\,hours every 28$\pm$3\,days. This adds up about 50\,hours per year ($\sim$0.6\% of the mission time).\\

No other \emph{Ariel} telescope operations are considered in the scheduler, either because they can be done simultaneously to other tasks, such as data downlink, or because their duration is very short (flat field images, dark exposures, etc.) and they can be easily scheduled during periods of inactive time (see Sect.\ref{subsec:resultsOther}).

\subsection{\emph{Ariel} list of targets}
\label{subsec:sample}
The list of targets that need to be observed is a key input of the scheduling algorithm. For the scheduler simulations we perform in this work we use a list adapted from ref.\,\cite{Edwards2019}, that describes the kind of targets that \emph{Ariel} can observe taking into account both the known planets up to now (more than 3300 transiting planets\footnote{http://exoplanet.eu}), and the predicted yield from ongoing surveys such as TESS (Transiting Exoplanet Survey Satellite, \cite{Ricker2015}). For each planetary system, the performance of the \emph{Ariel} instruments was estimated, resulting in $\sim$2000 planets that could be characterized. From this a Mission Reference Sample (hereafter, MRS) of 1000 planets was chosen (Edwards priv. comm.), including 384 known planets, and 616 predicted from the TESS yield \cite{Barclay2018}. This MRS is divided in three subsamples following the tier strategy approach of the \emph{Ariel} mission \cite{Tinetti2018}. These three subsamples are defined according to the spectral resolution and signal-to-noise ratio (SNR) that can be achieved from the observations:

\begin{itemize}
\item \underline{Tier 1 (T1)}. For all the targets in the MRS few transits or eclipses will be observed for each planet in order to get low-resolution spectra (several spectral bands from 1.1 to 7.8 $\mu$m) with SNR$\sim$7. Typically, for each target a sequence of up to 5 observations is requested.

\item \underline{Tier 2 (T2)}. A subsample of targets in the MRS will be observed to get higher spectral resolution ($\sim$10 below 1.9\,$\mu$m, $\sim$50 from 1.95 to 3.9\,$\mu$m, and $\sim$15 from 3.9 to 7.8\,$\mu$m) at SNR$\ge$7. The goal is to reach this level of precision for $\sim$600 targets. Up to 19 observations are needed for some of these stars.

\item \underline{Tier 3 (T3)}. About 50 interesting planets orbiting bright stars will be observed at the maximum spectral resolution of \emph{Ariel} ($\sim$15 below 1.9\,$\mu$m, $\sim$100 from 1.95 to 3.9\,$\mu$m, and $\sim$30 from 3.9 to 7.8\,$\mu$m) and SNR$\ge$7. Sequences of 1 or 2 observations are requested for each of these targets.
\end{itemize}

\begin{figure}[t]
\begin{center}
  \includegraphics[width=0.5\textwidth]{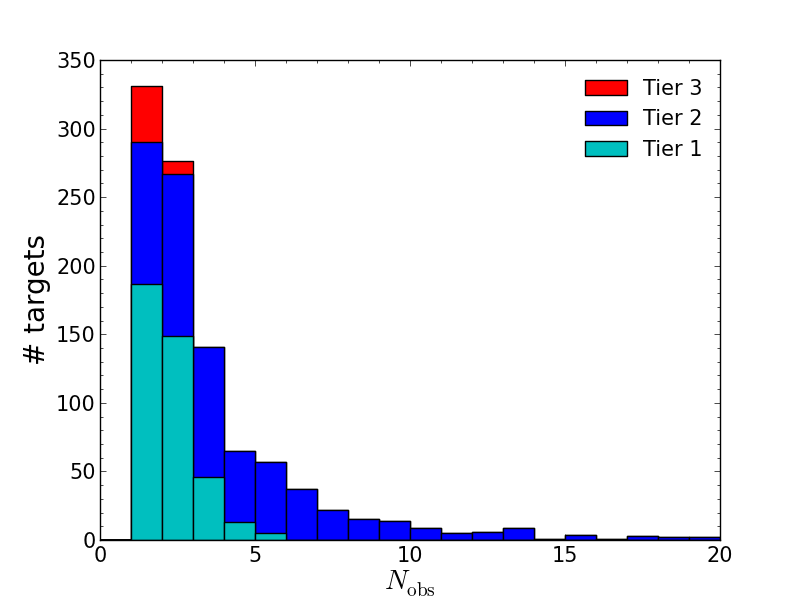}
\end{center}
\caption{Histogram of the number of targets as a function of the number of requested observations for each of them. Tier 1, tier 2, and tier 3 targets are plotted in different colors as labeled.}
\label{fig:nobs_T123}
\end{figure}

The key point for the scheduling exercise presented here is the number of observations and the amount of time needed for each subsample. Fig.\,\ref{fig:nobs_T123} illustrates the distribution of the number of observations for the different tier targets. The MRS used in this work, includes a total of 700 transits or occultations for the 400 exoplanets that will be observed only with tier 1 resolution. This accounts for $\sim$5650\,hours. 
For 550 stars, tier 2 resolution can be reached by following-up 2263 planetary events, which add a total of 15600\,hours of observing time.
Finally, to reach tier 3 level for 50 systems, a total of 59 observations lasting $\sim$330\,hours are needed.
In total, this adds up to $\sim$21600\,hours of exoplanet observations, which is $\sim70\%$ of the \emph{Ariel} nominal operations duration (18\%, 51\%, and 1\% for tiers 1, 2, and 3, respectively).

\begin{figure}[t]
  \includegraphics[width=0.5\textwidth]{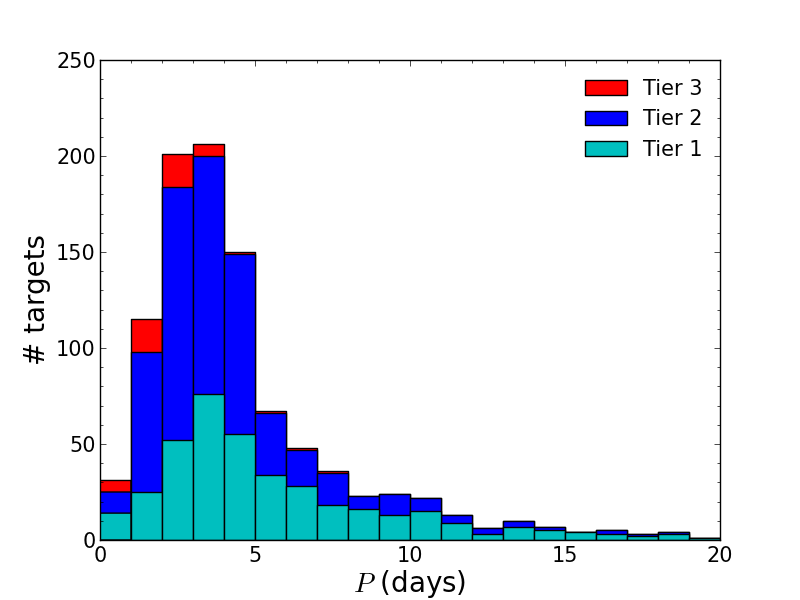}
  \includegraphics[width=0.5\textwidth]{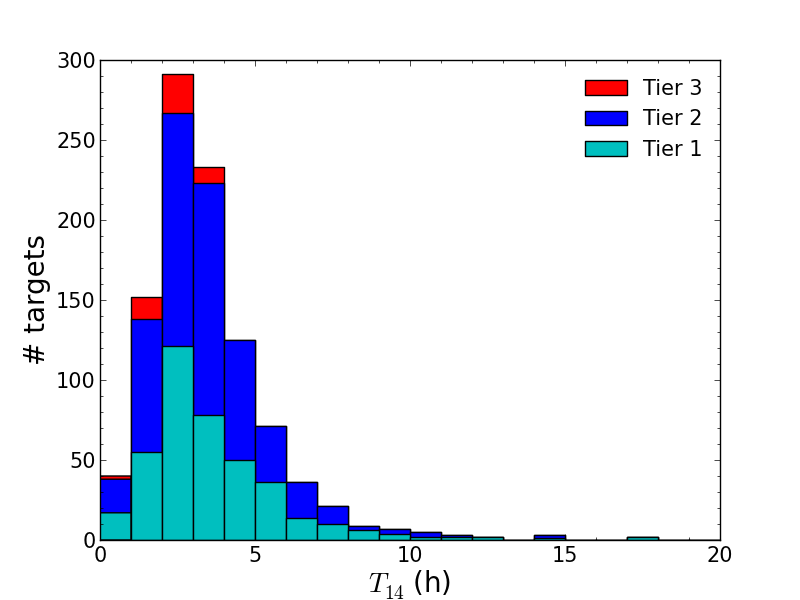}
\caption{Left: Histogram of the number of targets as a function of the exoplanet period. Right: Histogram of the number of targets as a function of the exoplanetary event duration. Tier 1, tier 2, and tier 3 targets are shown in different colors as labeled.}
\label{fig:period_T14_T123}
\end{figure}

A part from the tier and the number of observations, for each of the planets, the MRS provides the coordinates of its host star (right ascension and declination), the orbital period, the reference epoch for the transit and the occultation, the preferred type of event (transit or occultation) and the priority, which is based in the tier strategy approach, i.e., T3 targets have a higher priority than T2, and T2 than T1. Fig.\,\ref{fig:period_T14_T123} shows the distribution of the orbital period and the duration of the observable events for the targets in the MRS. Most of the exoplanets have periods between $\sim$0.5 and 30\,days, with a median value of $\sim$3.7\,days. Actually, only 9 systems (6 in T1 and 3 in T2 subsamples) have longer periods up to 324\,days. This means that most of the targets will have enough opportunities to schedule several transits or occultations. The duration of the planetary events, $T_{14}$, depends on the orbital period and the host star properties. It ranges from few tens of minutes to 18\,hours, but 90\% of the exoplanet targets have transits shorter than 6\,hours. The median duration of the MRS exoplanetary events is $\sim$3.5\,hours. 

Completing the observations of the MRS is the main goal of the \emph{Ariel} mission. However, two additional subsamples are provided to consider their observations:

\begin{itemize}
    \item \underline{Tier 4 (T4)}: for several short period systems phase curve observations would be desirable if not strongly affecting the mission core science. In order to test the feasibility of this subsample of observations, a list of 43 interesting systems distributed in three levels of priority were provided (B. Charnay priv. comm.). These are exoplanet systems with periods shorter than 7 days for which the continuous follow-up of 1 to 3 orbital cycles is requested.
    
    \item \underline{Back-up targets}: additionally a total of 1093 exoplanets were also provided as back-up targets in case some exoplanets in the MRS are not observable or cannot be completed during the mission lifetime. Their period is typically larger than for T1 targets in the MRS and they need to stack few more exoplanet events to reach T1 resolution. 
\end{itemize}

In this work, we use these subsamples to study different scenarios. From the one hand, we analyse the impact of including phase curves in the MRS by considering the exoplanets in the T4 subsample as the higher priority targets in the survey (see Sect.\,\ref{subsec:resultsPC}). On the other hand, we study if it would be possible to increase the number of exoplanets surveyed by \emph{Ariel} by including back-up targets in the plan with the same priority as the T1 subsample (see Sec.\,\ref{subsec:resultsBU}.)

\subsection{Mission constraints}
\label{subsec:constraints}
An automatic planning tool can be understood as an algorithm devoted to solve a constraint satisfaction problem, which is a mathematical problem defined as a set of objects whose states must satisfy a number of limitations. Some of these constraints need to be necessarily satisfied while other ones indicate what kind of solutions are preferred. These are typically identified as "hard" and "soft" constraints, respectively (see e.g. \cite{GarciaPiquer2015}).

\paragraph{\textbf{Hard constraints.}}
In the case of \emph{Ariel}, the hard constraints are mainly related to the visibility of the targets. They must be fulfilled to produce valid plans. We list here these constraints, which are obtained from the mission requirements documents.
\begin{itemize}
    \item \underline{Nominal operations duration}. \emph{Ariel} requirements state that the mission lifetime is 4 years. The first 6 months are foreseen for orbit and commissioning purposes. Therefore, with an expected launch in late 2028, nominal operations are expected to last 3.5 years starting in mid-2029. For the scheduling exercises presented here, we assume that exoplanet observations will run from July 1st 2029, to December 31st 2032.
    
    \item \underline{Orbital constraint}. The orbit and the attitude limits of the spacecraft must be considered by the scheduling algorithm because they are essential to determine when the targets are observable. According to the mission requirements, we assume that \emph{Ariel} will be positioned at the Sun-Earth Lagrangian point L2 and that the field of regard of the telescope is a cone in the perpendicular direction of the Sun-Earth direction with and angle between $+20^\circ$ and $–30^\circ$ toward and away of the Sun, respectively. Fig.\,\ref{fig:sample} displays the distribution of targets in right ascension and declination and the fraction of the year that each region of the sky is visible. Any direction is observable at least $\sim$30\% of the year.
    
\begin{figure}[t]
\begin{center}
  \includegraphics[width=0.75\textwidth]{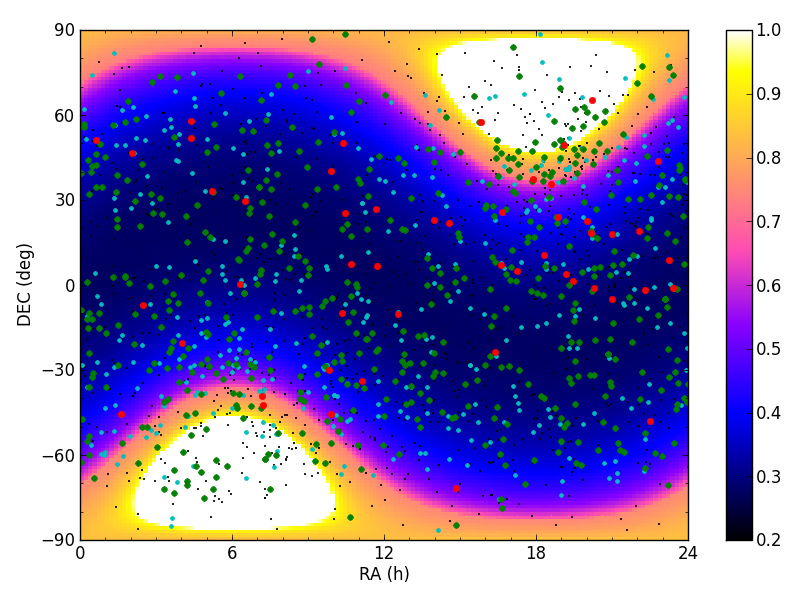}
\end{center}
\caption{Distribution of the \emph{Ariel} targets in the sky. Dot symbols correspond to tier 1 (cyan), tier 2 (green), tier 3 (red) and back-up targets (black). The background color levels indicate the fraction of the year for which each region of the sky is visible for \emph{Ariel}.}
\label{fig:sample}
\end{figure}

    \item \underline{Event timing and duration}. As described in Sect.\,\ref{subsec:operations}, \emph{Ariel} observational strategy consists in the observation of exoplanet transits, occultations, and phase curves, which are time-constrained according to the orbital ephemerides. The time at which each target can be observed is defined by the orbital period and a reference time, and the total duration of the follow-up is set to 2.5$\times T_{14}$ centered on the transit or occultation mid-time.
    
    \item \underline{Target completeness}. The MRS provides the number of transits or occultations that need to be stacked for each target in order to reach the required SNR. This defines a sequence of observations for each exoplanet and for each tier. Only completed sequences are useful in terms of science. For this reason, we added to the scheduler a constraint which removes from the planning all sequences that cannot be completed. Thus for any target, if T3 sequence cannot be completed, only T2 is considered, and if this is \textbf{also not} possible, only T1 sequence is planned. If T1 number of observations cannot be reached, then the target is removed from the scheduler. On the other hand, targets for which all required sequences have been planned, are no longer considered by the scheduler, unless revisits of targets are allowed.
    
    \item \underline{Slew of the telescope}. The movement of the telescope from one target to the next one should also be taken into account in the scheduler because it takes a significant fraction of the mission lifetime, and sufficient time should be left between target observations. According to \emph{Ariel} requirements, we assume a slew rate of 4.5\,deg/minute, and we add 5 minutes for telescope settling, reaction wheel off-loading and transition to fine pointing as indicated in the requirements.
    
\end{itemize}

\paragraph{\textbf{Soft constraints}}
Soft constraints are related \textbf{to} the optimization of the mission plan. They are not required to be fulfilled, but they are used to promote some solutions over the others. In the case of the \emph{Ariel} scheduler, at the present stage, we define two figures of merit that should be maximized:
\begin{itemize}
    \item \underline{Observing time}. Plans maximizing the total time used to observe exoplanet targets are preferred. \emph{Ariel} is expected to devote more than 85\% of the time to science observations, including the follow-up of exoplanets, calibration stars, and ancillary science.
    
    \item \underline{Number of completed targets}. The scheduler must also promote the number of targets in the MRS for which sequences are completed. The priority of each target is taken into account.
\end{itemize}
The optimization of these two soft constraints aims at obtaining mission plans with a large number of observations as well as a large number of completed targets.

\section{\emph{Ariel} scheduler}
\label{sec:algorithm}

The scheduling tool used for the simulations presented here, has been developed based on a detailed performance analysis of our previous studies on the \emph{Ariel} mission planning using EA \cite{GarciaPiquer2016,GarciaPiquer2017,GarciaPiquer2015}. The goals of the new development were first, to reduce the computational cost, and second, to increase the flexibility of the tool in order to adopt it to the updated mission constraints, simulation configuration and observation modes, such as phase curves. The new tool manages to obtain better results than its predecessor due to its flexibility to plan and optimize time for all the different tasks in a same process.

\begin{figure}[t]
\centering
  \includegraphics[width=8cm, height= 4.5cm]{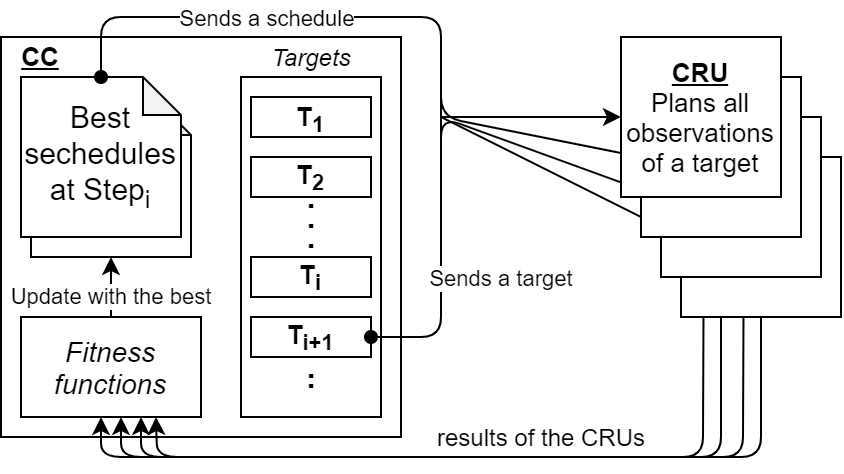}
\caption{Scheduling framework.}
\label{fig:DSP}
\end{figure}

The scheduling tool uses a Multi-start meta-heuristic hybrid algorithm (Nakhjiri et al. in prep.) developed around the problem specifications. The Algorithm has two main modules. First, is the lower-level Conflict Resolution Unit (CRU), which utilizes the flexibility of heuristic methods as its core approach, by following a Tabu-Search optimization \cite{glover1989tabu,glover1990tabu} in a neighborhood search \cite{yanez2003optimization}. Second, is the upper-level Control Center (CC), which manages several CRUs and has the fitness functions to evaluate the outcome of each CRU on the targets at every step. This allows us to search on numerous neighborhoods in the solution space and return the best overall findings as the final output. This process is illustrated in Figure \ref{fig:DSP}. 

In the rest of this section, first we explain the structure of the upper-level module, or CC, in \ref{subsec:cc}, and later we detail the process of lower-level module, or CRU, and its evaluation measures in \ref{subsec:cru}.

\subsection{Upper-level process}
\label{subsec:cc}
The upper-level of the algorithm or CC starts with an empty schedule and a full list of targets to plan at step 0 ($S_0 = \o $). At each step, it sends the current best schedule and a new target to all its CRUs. Each CRU plans the new observations one at a time in the best way it can and returns the result to CC. While a CRU always plan the tasks it is assigned to, CC has the job of calculating and evaluating the overall fitness of the CRU results and compares it with the previous step to decide whether to keep or discard the changes suggested by a CRU. Following a Swarm Intelligence framework, at each step CC gathers the results of its population of CRUs along with the best schedules of the previous step and based on the soft constraints of the problem, performs a elitist selection on their Pareto Optimal Front \cite{Pareto1897,Coello2001}. The selected solutions replace the list of best schedules in CC and will be used as a CRU input for the next step.

Almost all the hard-constraints of the ARIEL problem are handled prior to the scheduling process. The exceptions are the target completeness and the slew times of the telescope. While CRU considers the slew times, CC checks for the target completeness and removes the unwanted observations before checking it for the soft constraints. The soft-constraints considered for the simulations are first, to maximize the number of completed targets with regards to their priority ($F_C$) and second, to maximize the overall observation time of the survey ($F_O$). The fitness function $F_C$ for a schedule at step $i$ ($S_i$) is computed as 
\begin{equation}
\label{def:fc}
F_C(S_i) = \sum_{j=1}^i \vartheta * Pri(t_i)
\end{equation}
where $Pri(t_i)$ refers to the priority of the target $t_i$ and $\vartheta$ is the SNR satisfaction level multiplier. According to the number of planned observations and the reached SNR levels, $\vartheta$ increases by each SNR level reached. The increment is doubled for the first SNR level in order to encourage entry of new targets to the schedule. The second fitness function, $F_O$, is measured as 
\begin{equation}
\label{def:fo}
F_O(S_i) = \sum_{O \in S_i} |O|
\end{equation}
where $O$ refers to the planned observations in $S_i$, and $|O|$ represent the duration of $O$. These fitness functions are used to evaluate the outcome of CRUs. They are normalized to the total duration of the \emph{Ariel} nominal operations and the total number of targets, and an equivalent weight is given for these figures of merit, providing a single solution, instead of a Pareto optimal set of solutions.

\subsection{Lower-level process}
\label{subsec:cru}
\begin{figure}[t]
\centering
  \includegraphics[width=\textwidth]{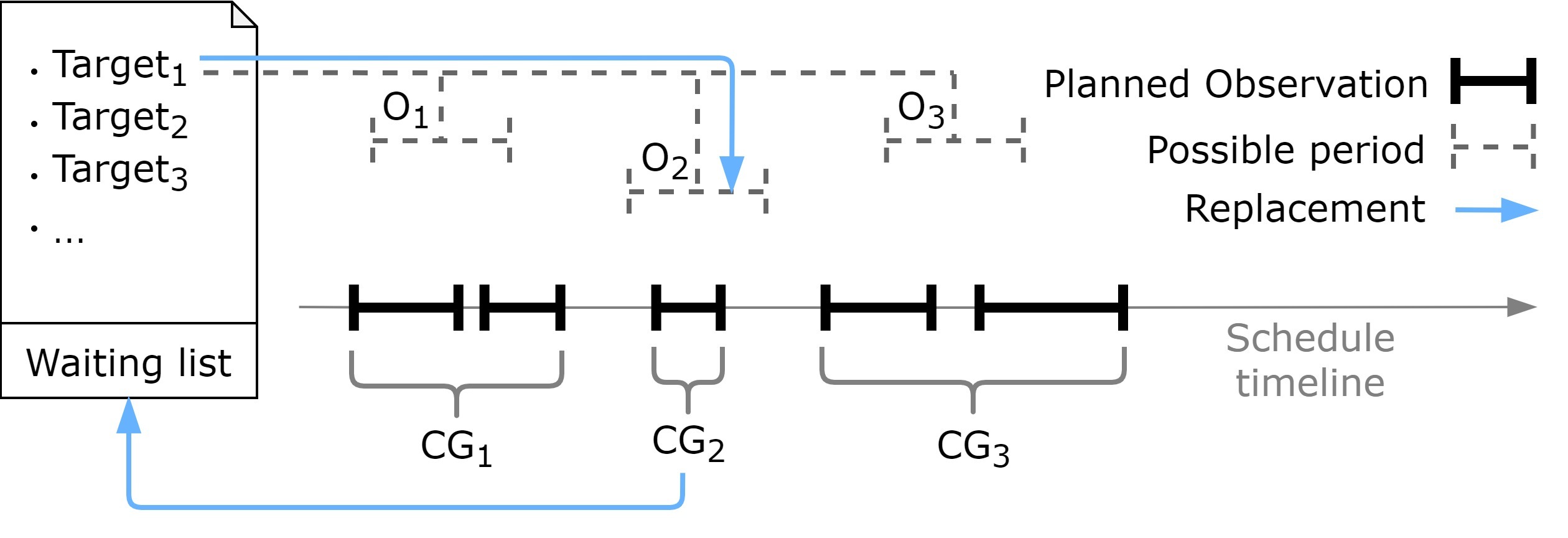}
\caption{CRU cycle. The replacement marks illustrate that Target$_{1}$ is removed from the waiting list and it is added to the schedule in the possible period O$_{2}$. Targets in CG$_{2}$, which is the Conflict Group of O$_{2}$ are removed from the schedule and are added to the waiting list.}
\label{fig:CRU}
\end{figure}

In the lower-level part of the algorithm or CRU, new observations are always planned regardless of their overall cost. The cost of planning for an observation is measured based on the priority of the tasks it has to replace in order to fit into the schedule, as 
\begin{equation}
\label{def:cost}
\sigma \in CG(O) \Leftrightarrow \sigma \,\, conflicts\, with\,\, 
O
\end{equation}
\begin{equation}
\label{def:cost2}
C_O = \sum_{\sigma \in CG(O)} \Delta \vartheta * Pri(\sigma)\\
\end{equation}
where $CG(O)$ is the Conflict Group of observation $O$ and contains all the tasks, $\sigma$, in the schedule that overlap with it. Also, $\Delta \vartheta$, shows the difference in the $\vartheta$ multiplier when losing an observation of a target. Cost is defined as the aggregated $F_C$ loss of all conflicts. In this format, $C_O = 0$ indicates a free spot for an observation. CRU job is to minimize this cost for each observation that it is assigned to schedule. After receiving a target, CRU plans for its observations one at a time in a recursive cycle. The process is triggered by adding a new observation to the waiting list ($L$) of CRU. It is then replaced by a $CG$ of a low cost. The new observation fits into the schedule and the members of the $CG$ are moved to $L$. On the next iteration, another member of $L$ is selected and replaced. This process is illustrated in Figure \ref{fig:CRU}.

The cycles continue until $L$ becomes empty, meaning all the observations have been fitted into the plan, or CRU reaches one of its limits. The limits are first, reaching the maximum number of iterations, and second, there is no new option to replace as it falls into a loop. To choose a replacement, the selection is based on the cost of each possible period in a way that the lower the cost is, the higher the chance of getting replaced. The odds of replacement for an observation $O$, $R(O)$, where $O$ belongs to the set of observation periods of a target $t$ ($P_t$), is calculated as 
\begin{equation}
\label{def:cost}
R(O) = \frac{\sfrac{1}{C_O}}{\sum_{\sigma \in P_t} \sfrac{1}{C_{\sigma}} }
\end{equation}
This weighted random selection allows us to explore the more relevant parts of the solution space by different CRUs.
After processing all the observations of a new target, CRU returns the best finding for each of them to CC and let it decide to keep or discard them. In this way, even if
planning the sequence of a target damages the current schedule more than it contributes to its competence, we can see the impact.

The abstract schema for the tasks handling in the method, allows us to define and process any type of tasks, including science observations, calibration, house keeping and slews in the same format and place. This results in more coherent optimization between all the pieces. Also, the step-scheduling framework, gives us more flexibility to setup different tests and assess various simulation configurations. These improvements result in a significant increase in the number of completed targets of ~2.5\% and 3.3\% for T1 and T2 targets, respectively, with respect to the \emph{Ariel} scheduler code based on MOEA presented in ref. \cite{GarciaPiquer2015} using the MRS. As a consequence the time used for science increases by $\sim$5.5\%. This improvement is mainly due to a reduction in the number of necessary calibrations by $\sim$10\% although still fulfilling the expected cadence (see Sect.\,\ref{subsec:resultsMRS}).
The reduction was possible with coherent optimization of operation tasks along side the scientific ones, in the new tool, instead of planning these tasks in different steps as in the former version.

The scheduler tool presented here is developed in C++ language and the execution of the experiments is done on a desktop computer with 6 cores at 2.2 GHz and 8 GB of RAM. This scheduler is able to produce an optimized plan for the 3.5 years of the \emph{Ariel} nominal operations phase in between 20 seconds to 10 minutes depending on the target set and imposed constraints. For example, scheduling revisits to targets takes longer computations. This will allow to regularly re-compute an optimized plan taking into account former observations, updates to the list of targets, or targets of opportunity that may arise during the mission, as part of the activities of the \emph{Ariel} Ground Segment and Instrument Operations Science Data Centre \cite{Pearson2020}.

\section{\emph{Ariel} mission planning simulations}
\label{sec:results}

Scheduler tools are not only useful to plan observations but also to simulate the outcome of the mission. As discussed above, the main science goal of \emph{Ariel} is the characterization of the atmosphere of $\sim$1000 exoplanets through transit and occultation spectroscopy. To evaluate if this is feasible within the 3.5\,years of nominal mission operations, we have made use of the scheduler described above to simulate the planning of the targets in the MRS taking into account all the mission constraints. It is also possible to test different mission scenarios with such simulations. For instance, \emph{Ariel} may be able to study the phase curve of several targets. To do so, it would need to follow-up the same target for a much longer time than for a transit or an occultation. Therefore, they could have a strong impact on the number of MRS targets that can be completed. Besides, it is also possible to estimate what is the maximum number of targets that \emph{Ariel} could be able to observe.

We explore the scheduling of these different scenarios in the following sections. In all the cases, the simulations performed here fulfill all the hard mission constraints listed in Sect.\,\ref{subsec:constraints}. Besides, we remind that we consider the optimization of the soft constraints, which are \emph{1)} the number of completed targets, and \emph{2)} the telescope working time.

\subsection{Scheduling the MRS targets}
\label{subsec:resultsMRS}

In this case, we have only considered the scheduling of the 1000 targets in the MRS. It includes a total of 3022 events lasting about 21600 hours. This number of hours is about 70\% of the time available during nominal operations (3.5 years); therefore, in principle, all observations should fit in the mission plan. However, since observations are time constrained according to the ephemerides, and the sequence of each target can include several transits or occultations, overlapping between simultaneous events could prevent the completion of some targets.

Table\,\ref{tab:planetsMRS} shows the number of planets whose observational sequences can be completed within the 3.5\,years as a function of the type of exoplanet according to its size and temperature (see \cite{Edwards2019} for details). Almost all targets can be scheduled except 2 temperate Jupiter-like planets, which correspond to exoplanets HIP\,41378f \cite{Vandenburg2016} and PH-2b \cite{Wang2013}, both in the T1 subsample. Actually, HIP\,41378f is not schedulable because due to its long period reported, 324 days, only 1 transit is observable while 2 are requested to reach tier 1 resolution level. On the other hand, PH-2b is also a system with a long period, 282.5 days, for which all the 4 transits that \emph{Ariel} could observe must be scheduled. Each transit observation takes about 1.2 days of continuous follow-up. This long duration increases the probability of overlapping with other higher priority targets, and it is finally discarded by the scheduler.

\begin{table}[t]
\centering
\caption{Number of exoplanets included in the mission plan scheduling simulation over the total number of targets in the MRS (in bold).}
\label{tab:planetsMRS}
\begin{tabular}{cccccc}
\hline\noalign{\smallskip}
Planet type$^1$      & Ultra Hot & Very Hot & Hot & Warm & Temperate  \\
\hline\noalign{\smallskip}
Massive Jupiter  & 8/\textbf{8}   & 115/\textbf{115} & 10/\textbf{10}   & -- & -- \\
Jupiter          & 43/\textbf{43} & 300/\textbf{300} & 234/\textbf{234} & 79/\textbf{79} & 5/\textbf{7}          \\
Neptune          & 4/\textbf{4}   & 14/\textbf{14}   & 21/\textbf{21}   & 26/\textbf{26} & 4/\textbf{4} \\
Sub-Neptune      & 1/\textbf{1}   & 8/\textbf{8}     & 14/\textbf{14}   & 39/\textbf{39} & 27/\textbf{27} \\
Super-Earth      & --             & 2/\textbf{2}     & 6/\textbf{6}     & 18/\textbf{18} & 20/\textbf{20} \\
\hline\noalign{\smallskip}
\end{tabular}
\flushleft
$^1$Planet type corresponds to the definition provided in the target list \cite{Edwards2019}.
\end{table}

In summary, a total of 998 targets can be scheduled within the \emph{Ariel} nominal operations phase, reaching T3 and T2 spectral resolution for 50 and 550 targets respectively, fulfilling science goals. With respect to the time usage, first row in Table\,\ref{tab:resultsMRS} lists the total number of observations and the time used for each spacecraft operation. A total of 3016 planetary events are scheduled. Target observations take about 21400 hours or $\sim70$\% of the 3.5 years. About 1100 hours are spent slewing between targets (including stabilization time). Finally, 943 and 172 hours are used for calibrations and station keeping, respectively, fulfilling the expected cadence, the median separation between these operations is 1.8 days and 30 days, respectively.

\begin{table}[t]
\centering
\caption{Number of completed targets, total number of observations, and distribution of time between the different tasks corresponding to the scheduling simulation of the MRS.
}
\label{tab:resultsMRS}       
\begin{tabular}{lccccc}
\hline\noalign{\smallskip}
              &          &              & \multicolumn{3}{c}{Total time (hours / \%$^1$)} \\
Case          & Targets  & Observations & On targets & Slewing & Waiting time \\
\noalign{\smallskip}\hline\noalign{\smallskip}
MRS           & 998      & 3016         & 21376.9    & 1101.9  & 7126.2 \\
              &          &              & 69.59\%    & 3.59\%  & 23.20\%\\
MRS-fill      & 998      & 3653         & 23391.2    & 1332.7  & 4881.1 \\
              &          &              & 76.14\%    & 4.34\%  & 15.89\%\\
\noalign{\smallskip}\hline
\end{tabular}
\flushleft
$^1$Fraction of time with respect to the duration of the \emph{Ariel} nominal operations phase (3.5 years).
\end{table}

\begin{figure}[t]
  \includegraphics[width=0.5\textwidth]{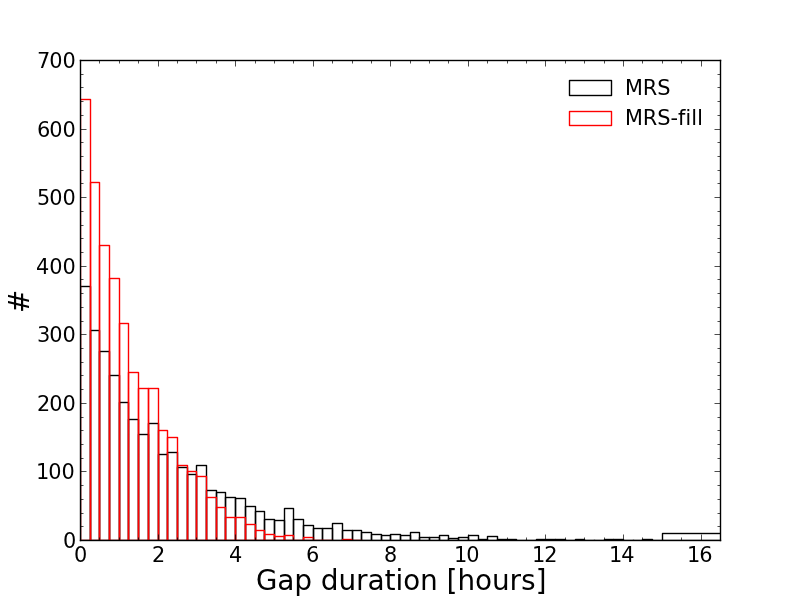}
  \includegraphics[width=0.5\textwidth]{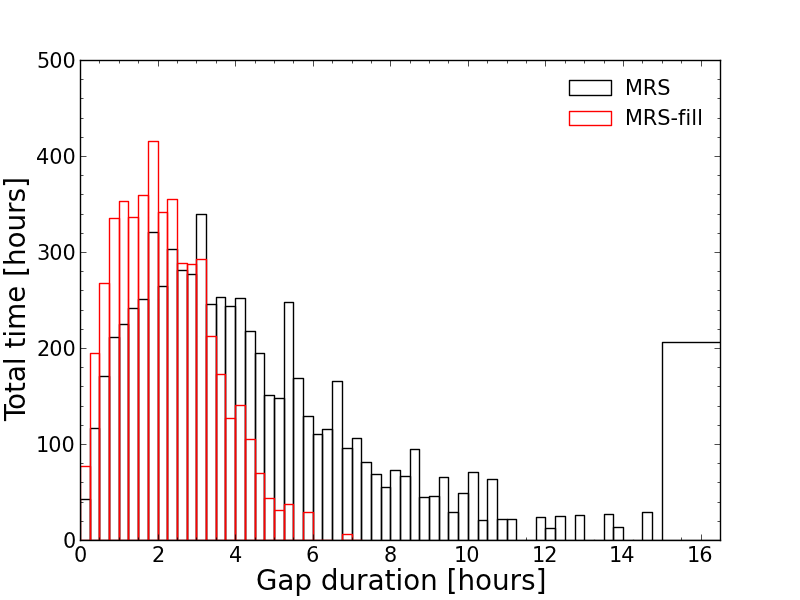}
\caption{Left: Histogram of the number of inactive periods between observations, gaps, as a function of their duration. Right: Distribution of the cumulative time for each gap duration. Black and red lines correspond to simulations of the MRS and the MRS including revisits to targets, respectively. Last bin corresponds to few inactive slots larger than 15\,hours.}
\label{fig:gaps_statistics}
\end{figure}

The last column in Table\,\ref{tab:resultsMRS} lists the waiting time that remains between target observations, which is neither used for any of the other tasks included in the scheduler.
This is an inevitable outcome of \emph{1)} scheduling time constrained events, and \emph{2)} only scheduling the number of observations for each target sequence given in the target list. Hereafter, we refer to these periods of unused waiting time as "gaps". The panels in Figure\,\ref{fig:gaps_statistics} show the distribution of the number of gaps and their total time (black line). Most of them are shorter than 1.5\,hours, but they accumulate a small number of hours. Longer gaps are smaller in number but they accumulate a significant fraction of time. Actually, these long gaps can be used to schedule further observations of targets. In total, they add up to 23.2\% of the \emph{Ariel} nominal operations duration.

Although revisits to targets are not a requirement, we have studied if further MRS transits or occulations could be fitted within these gaps. To do so, we allowed the scheduler to plan more observations than requested for each target whenever possible without interfering with completed targets.
Second row in Table\,\ref{tab:resultsMRS}, labeled "MRS-fill", summarizes the results of this simulation. A total of 637 additional events can be planned, adding $\sim$2000\,hours to target observations. This provides further observations for each target and flexibility to re-schedule failed observations due to unexpected problems. The amount of added time is about 10\% of that needed to complete the MRS.
Obviously, such revisits increase significantly the time devoted to slew between targets, but waiting time in gaps between observations is reduced to $\sim$4900 hours. The number and cumulative duration of gaps for this simulation are represented as red bars in Figure\,\ref{fig:gaps_statistics}. No additional exoplanet transits or occultations for MRS targets can be fitted within these gaps. Actually, most of them are much shorter than before. Their median and maximum duration are $\sim$1\,hour and 6.7\,hours, respectively. 

\begin{figure}[t]
  \includegraphics[width=\textwidth]{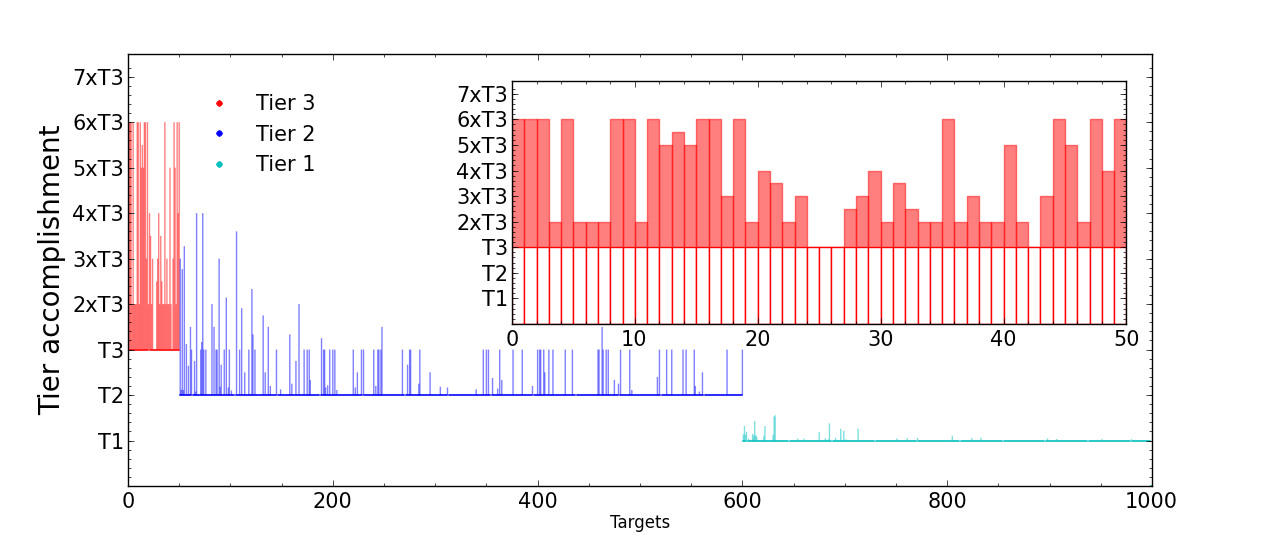}
\caption{Tier accomplishment for each planetary system in the MRS. The bottom and top values of each vertical line depicts the observation level for each target for simulations excluding and including refilling of gaps, respectively. Targets are sorted according to their classification in tier 1, tier 2, and tier 3, in increasing order of the exoplanet orbital period (from left to right), \textbf{and plotted in different colors as labeled}. Inset plot displays the case of tier 3 targets using bars. Decimal values in the Y-axis indicate the fraction of observations to reach the next tier level.}
\label{fig:planet_visits}
\end{figure}

These additional observations might be extremely useful to study the time variability of exoplanet atmospheres, or to increase the SNR for some targets by piling-up a larger number of transit or occultations. Figure\,\ref{fig:planet_visits} depicts the improvement reached for each target when allowing revisits. Y-axis on this figure indicates the level of accomplishment of each target as a function of its tier, while the X-axis indicates the number of target, sorted by tier subsample and orbital period. Additional sequences are more easily scheduled for short period systems because they are shorter and the number of opportunities is longer. A total of 19 T3 sequences can be re-observed at least 5 more times, thus allowing to study exoplanet variability. Besides, 83 T2 targets can be promoted to T3, potentially increasing the number of exoplanets studied at the highest resolution and SNR level to 133 exoplanets. Finally, several T2 and T1 targets can be revisited although not reaching T3 or T2 levels.

As a final conclusion, about 23400\,hours can be used to schedule $\sim$3650 planetary transits and occultations of almost all the targets in the MRS, reaching the requested SNR level and potentially improving it for several targets.

\subsection{Impact of scheduling phase curves}
\label{subsec:resultsPC}
To study the impact of including the follow-up of several phase curves in the \emph{Ariel} mission planning, we have applied our scheduler algorithm to the MRS targets including the 43 exoplanet systems in the T4 subsample for which phase curves would be observed, if possible, as if they were independent observations. For this trade-off analysis, we have considered three different cases that progressively take into account the three priorities in which the T4 target list is distributed. In "PC-case1" we only consider the scheduling of the 22 highest priority T4 exoplanets, in "PC-case2" we add 15 medium priority T4 targets, and finally "PC-case3" includes all the 43 T4 exoplanet systems. The total exposure time needed to follow-up phase curves amounts to 4.6\%, 9.9\%, and 11.6\% of the \emph{Ariel} nominal operations duration for PC-case1, PC-case2, and PC-case3, respectively. Although this fraction of time could still be fitted within the mission lifetime, our simulations show that it is not possible to schedule the observations of all MRS and T4 targets. This is mainly due to the overlap between observations caused by the long duration of the phase curve follow-up, which range between $\sim$1 to $\sim$9\,days.

Table\,\ref{tab:resultsPC} lists the results of the mission plans obtained for each studied case. For these simulations, we made sure to plan all T4 targets by setting their priority above that of T3 exoplanets in the MRS. Phase curve observations reduce the number of MRS completed targets by 1.4\%, 5.8\%, and 6.9\% for cases 1, 2, and 3, respectively. The most affected subsample of the MRS is tier 1 due to its lower priority, missing 14, 58, and 69 T1 targets in each case. Only 2, 9, and 18 T2 targets are downgraded to tier 1 SNR, for cases 1, 2, and 3, respectively, while T3 target requirements are always fulfilled. The number of exoplanets and phase curves included in the different mission planning simulations for each kind of planet system is provided in Table \ref{tab:planetsPC}. In particular, the number of surveyed Neptune to Earth-like planets in the MRS would be significantly reduced by $\sim$5--16\%.

\begin{table}[t]
\centering
\caption{Number of completed MRS and T4 targets, total number of observations, and distribution of time between the different tasks corresponding to the scheduling simulation of the MRS and phase curves.}
\label{tab:resultsPC}       
\begin{tabular}{lccccc}
\hline\noalign{\smallskip}
              & Targets  &              & \multicolumn{3}{c}{Total time (hours / \%$^1$)} \\
Case          & MRS + T4 & Observations & On targets & Slewing & Waiting time \\
\noalign{\smallskip}\hline\noalign{\smallskip}
PC-case1      & 984 + 22 & 3021         & 22521.2    & 1087.6  & 6000.2 \\
              &          &              & 73.31\%    & 3.54\%  & 19.53\%\\
PC-case2      & 940 + 37 & 2913         & 23185.9    & 1048.4  & 5403.7 \\
              &          &              & 75.47\%    & 3.41\%  & 17.59\%\\
PC-case3      & 929 + 43 & 2834         & 23278.4    & 1017.2  & 5345.4 \\
              &          &              & 75.78\%    & 3.31\%  & 17.40\%\\
\noalign{\smallskip}\hline         
PC-case1-fill & 984 + 22 & 3364         & 23488.9    & 1201.7  & 4918.4 \\
              &          &              & 76.46\%    & 3.91\%  & 16.01\%\\
PC-case2-fill & 940 + 37 & 3185         & 23884.7    & 1146.6  & 4606.7 \\
              &          &              & 77.75\%    & 3.73\%  & 15.00\%\\
PC-case3-fill & 929 + 43 & 3137         & 23930.3    & 1111.6  & 4599.1 \\
              &          &              & 77.90\%    & 3.62\%  & 14.97\%\\
\noalign{\smallskip}\hline
\end{tabular}
\flushleft
$^1$Fraction of time with respect to the duration of the \emph{Ariel} nominal operations phase (3.5 years).
\end{table}

\begin{table}[t]
\centering
\caption{Number of exoplanets included in the mission plan scheduling simulation of the MRS and phase curves (T4).}
\label{tab:planetsPC}
\begin{tabular}{cccccccccc}
\hline\noalign{\smallskip}
                & \multicolumn{2}{c}{Ultra Hot} & \multicolumn{2}{c}{Very Hot} & \multicolumn{2}{c}{Hot} & \multicolumn{2}{c}{Warm} & \multicolumn{1}{c}{Temperate$^3$}  \\
Planet type$^1$  & MRS       & T4$^2$ & MRS         & T4$^2$ & MRS          & T4$^2$ & MRS      & T4$^2$  & MRS       \\
\hline\noalign{\smallskip}
Massive Jupiter  & 8,8,8     & 1,1,1  & 113,112,113 & 5,7,10 & 10,10,10     & --     & --        & --     & --       \\
Jupiter          & 43,39,40  & 1,1,1  & 298,291,284 & 5,8,10 & 234,229,223  & 3,6,6  & 79,76,74  & --     & 4,4,4    \\
Neptune          & 4,2,2     & --     & 14,10,10    & 1,1,1  & 21,19,20     & 1,3,3  & 26,24,22  & 1,2,2  & 2,2,2    \\
Sub-Neptune      & 1,0,1     & --     & 8,5,7       & 2,4,5  & 12,11,11     & 0,2,2  & 35,35,29  & 1,1,1  & 27,25,27 \\
Super-Earth      & --        & --     & 2,1,1       & --     & 5,3,3        & --     & 18,16,18  & 1,1,1  & 20,18,20 \\
\hline\noalign{\smallskip}
\end{tabular}
\flushleft
$^1$Planet type corresponds to the definition provided in the target list \cite{Edwards2019}.

$^2$The three values indicated for each case correspond to the simulations of PC-case1, PC-case2, and PC-case3 (see text for further details).

$^3$T4 sample does not include temperate planets.
\end{table}

On the other hand, the scheduling of phase curves helps to optimize the total time used for exoplanet observations. As indicated in Table\,\ref{tab:resultsPC}, the total time tracking exoplanets increases with respect to the solution in Sect.\ref{subsec:resultsMRS} by about 5.4\%, 8.5\% and 8.9\% in PC-case1, PC-case2 and PC-case3, respectively, despite scheduling a smaller number of MRS targets. This is due to the longer duration of the follow-up of phase curves with respect to transits and occultations of MRS targets. As a consequence, the waiting time is also reduced, by $\sim$16--25\%. Actually, in the most critical case depicted by PC-case3, the time needed to fit the phase curves in the plan is approximately taken half from the omitted MRS targets and half from gaps. Only a small fraction comes from the reduction of the total slew between targets. Finally, part of this time is also taken from the few short calibrations that cannot be planned fulfilling the expected cadence (see Sect\,\ref{subsec:constraints}). Short calibrations overlapping with long phase curves are moved before and after each T4 target observation. This reduces the number of short calibrations in 4, 33 and 36 cases, for PC-case1, PC-case2 and PC-case3, respectively.

As a further check, we run the mission planning simulations including the MRS and T4 planets allowing for the revisit of targets. Results are also provided in the second block in Table\,\ref{tab:resultsPC} labelled as "-fill". The number of additional observations is reduced by a factor of $\sim$2 with respect to the mission planning simulations of the MRS (Sect.\,\ref{subsec:resultsMRS}), and the total total time revisiting targets, by a factor $\sim$2--3. An inevitable outcome is that in this case, the number of targets that can be promoted from T2 to T3 SNR level is reduced to $\sim$50, and that T3 sequences can be re-observed only two times for 3--4 T3 targets.

Thus, while phase curve observations helps to maximize the time used for science, they reduce the number of MRS targets completed during the \emph{Ariel} mission lifetime. However,  it is important to note here that the follow-up of phase curves comprise at least 1 transit and 2 occultations, which might still put constraints on the exoplanet atmosphere composition. Furthermore, phase curves on their own will also be useful to explore the atmosphere circulation. The scheduling tool might be very helpful to find a balance between the MRS and phase curves targets.

\subsection{Mission planning optimization limits}
\label{subsec:resultsBU}
The results in Sect.\,\ref{subsec:resultsMRS} and \ref{subsec:resultsPC} indicate that, due to the time constrained nature of \emph{Ariel} observations, about 70--78\% of the nominal operations phase can be used for actual exoplanet observations. However, this number depends also on the sample of targets considered. Some flexibility on the targets chosen allows to select that ones that best fit in the mission plan. In order to analyze this fact, we have used our scheduler to compute mission plans including the MRS (T1, T2, and T3) and back-up (T0) targets. For this purpose, exoplanets in the T0 subsample were assigned a priority equal to that of T1 targets. The observations of all back-up targets up to the T1 expected SNR level adds up to $\sim$40000\,hours, which almost double the time needed to complete the MRS. This increases the telescope oversubscription factor by $\sim$2. Obviously, it is not possible to schedule all these targets, but this test provides an estimation of the total time that can be used to follow-up exoplanet events and the maximum number of targets that can be surveyed using a longer target list.

Table\,\ref{tab:planetsBU} lists the number of completed planets over the total number of targets in the input sample as a function of the planetary type. The factor increase with respect to Table\,\ref{tab:planetsMRS} is also provided for each value. The main difference is that the number of scheduled Neptune and sub-Neptune size planets increases by a factor of $\sim$3 and $\sim$2, respectively. Besides, a total of 12 warm and temperate Earth-like planets are added in the survey. This increase in the number of surveyed planets is mainly due to several reasons: \emph{1)} all back-up targets need few observations to reach T1 resolution, \emph{2)} the scheduler can choose between more targets to select those that best fit in the planning, \emph{3)} T0 targets have a larger population of Neptune to Earth-size planets. Actually, the number of exoplanets in the T0 sample is about 1.9, 5.4, 3, and 1.4 times that of the MRS for Jupiter, sub-Neptune, Neptune and super-Earth planets, respectively, while only 4 new massive Jupiter planets are added. As a consequence, the scheduling tool finds more efficient solutions by pushing the number of completed planets. This causes that in this scenario many of the exoplanets are observed only up to T1 resolution. Actually, from the 1486 targets planned, 50 are observed up to T3 level, 301 up to T2, and the rest up to T1.

\begin{table}[t]
\centering
\caption{Number of exoplanets included in the mission plan of the MRS and back-up targets (T0) over the total number of T0 to T3 targets, and increase factor$^1$ with respect Table\,\ref{tab:planetsMRS}.}
\label{tab:planetsBU}
\begin{tabular}{cccccc}
\hline\noalign{\smallskip}
Planet type$^2$      & Ultra Hot & Very Hot & Hot & Warm & Temperate  \\
\hline\noalign{\smallskip}
Massive Jupiter  & 8/\textbf{8}   & 112/\textbf{118}  & 11/\textbf{11}   & --               & --             \\
                 & 1.0/1.0        & 0.97/1.03         & 1.10/1.10        & --/--            & --/--          \\
Jupiter          & 46/\textbf{60} & 378/\textbf{507}  & 357/\textbf{485} & 137/\textbf{193} & 5/\textbf{11}  \\
                 & 1.07/1.40      & 1.26/1.69         & 1.53/1.07        & 1.58/2.44        & 1.0/1.86       \\
Neptune          & 4/\textbf{11}  & 37/\textbf{73}    & 77/\textbf{136}  & 86/\textbf{145}  & 4/\textbf{5}   \\
                 & 1/2.75         & 2.64/5.21         & 3.67/6.48        & 3.31/5.58        & 1.0/1.25.      \\
Sub-Neptune      & 1.0/\textbf{1} & 18/\textbf{24}    & 24/\textbf{51}   & 80/\textbf{136}  & 43/\textbf{55} \\
                 & 1.0/1.0        & 2.25/3.0          & 1.71/3.64        & 2.05/3.49        & 1.59/2.04.     \\
Super-Earth      & --             & 2/\textbf{2}      & 5/\textbf{6}     & 26/\textbf{27}   & 25/\textbf{28} \\
                 & --/--          & 1.0/1.0           & 0.83/1.0       & 1.44/1.50        & 1.25/1.40      \\
\hline\noalign{\smallskip}
\end{tabular}
\flushleft
$^1$The increase factors of the planned and total targets with respect to the mission planning simulation of the MRS presented in Table\,\ref{tab:planetsMRS} are given in the second row of each planet type.

$^2$Planet type corresponds to the definition provided in the target list \cite{Edwards2019}.
\end{table}

In terms of time usage, there are two significant differences with respect to the simulations described in Sect.\,\ref{subsec:resultsMRS}. Table\,\ref{tab:resultsBU} provides the distribution of time when including back-up targets between the different spacecraft operations. Adding more targets in the scheduler increases the time used for scientific observations by $\sim$2400\,hours. This results into a smaller fraction of time distributed in waiting times between observations. If revisits to targets are allowed to fill all possible gaps, 59 targets can be promoted from T2 to T3 and 5 T3 sequences can be repeated at least 5 times. Obviously these numbers are smaller than in the case of the MRS simulations (see Sect.\,\ref{subsec:resultsMRS}) because most of the waiting time is dedicated to complete other targets. This simulation refilling all of the possible gaps with exoplanet transits and occultations provides an estimate of the maximum time that can be used for exoplanet science given the current list of targets, which amounts to about 80\% of the \emph{Ariel} mission lifetime, $\sim$24600\,hours.

\begin{table}[t]
\centering
\caption{Number of completed targets, total number of observations, and distribution of time between the different tasks corresponding to the scheduling simulation of the MRS including back-up targets.}
\label{tab:resultsBU}       
\begin{tabular}{lccccc}
\hline\noalign{\smallskip}
              &          &              & \multicolumn{3}{c}{Total time (hours / \%$^1$)} \\
Case          & Targets  & Observations & On targets & Slewing & Waiting time \\
\noalign{\smallskip}\hline\noalign{\smallskip}
B/U        & 1486        & 3016         & 23804.2    & 1083.0   & 4702.8 \\
              &          &              & 77.49\%    & 3.52\%   & 15.31\%\\
B/U-fill   & 1486        & 3348         & 24576.9    & 1208.5   & 3804.6 \\
              &          &              & 80.0\%     & 3.93\%   & 12.38\%\\
\noalign{\smallskip}\hline
\end{tabular}
\flushleft
$^1$Fraction of time with respect to the duration of the \emph{Ariel} nominal operations phase (3.5 years).
\end{table}

\subsection{Use of inactive periods}
\label{subsec:resultsOther}

As already mentioned, a common outcome of the different simulations described above is that part of the time is lost in gaps of waiting time between observations. In the nominal case including only the MRS exoplanets in the scheduler, these gaps add up $\sim$7000\,hours and each of them could take several hours. The scheduling simulations presented above demonstrate that this time is reduced by a factor of $\sim$2 if exoplanet sequences are repeated, or phase curves or back-up targets are added in the plan.
This would significantly increase the scientific results of the mission, by adding more targets to the survey, increasing those observed at the highest resolution (T3), or studying exoplanet atmosphere variability, or atmospheric circulation. After this refilling of gaps, the optimized plans show that there are still about 4000-5000 hours distributed in gaps shorter than $\sim$7\,hours where no more transits or occultations can be planned (see Fig.\,\ref{fig:gaps_statistics}).

\begin{figure}[t]
  \includegraphics[width=0.5\textwidth]{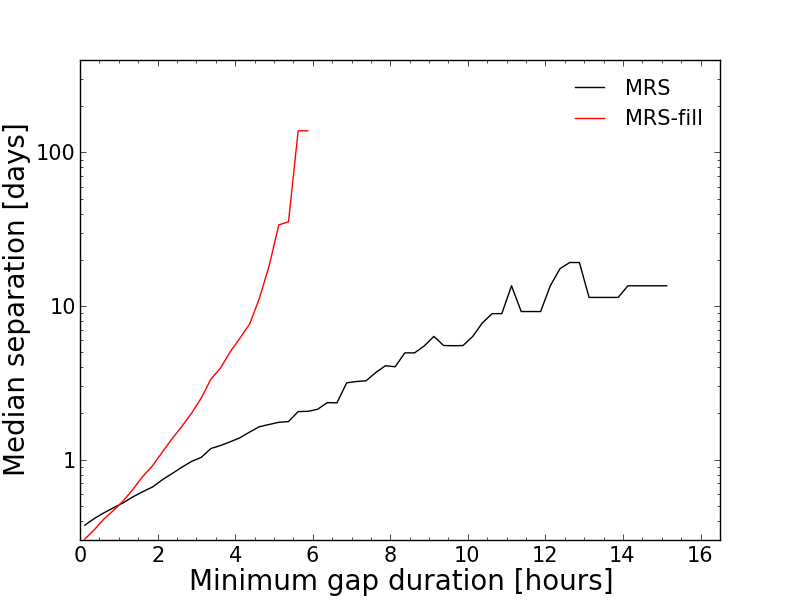}
  \includegraphics[width=0.5\textwidth]{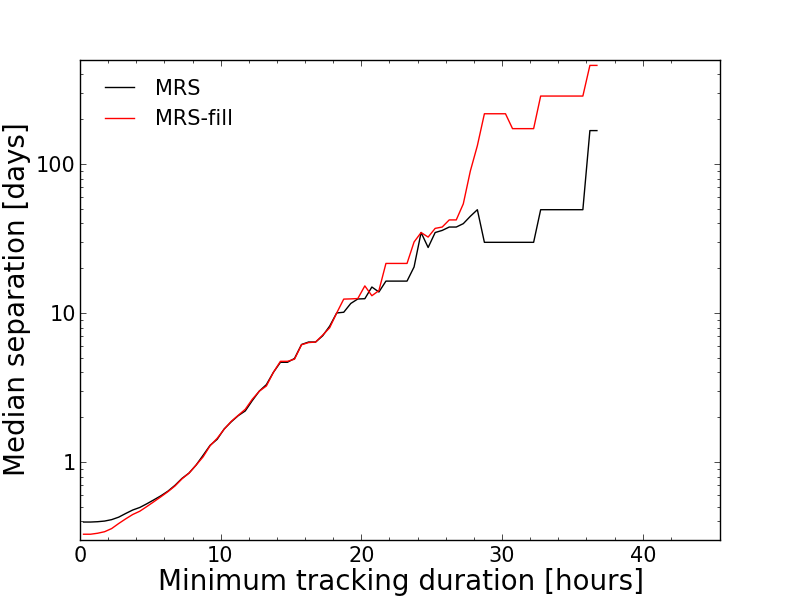}
\caption{Median separation between gaps (left) and tracking on targets periods (right) as a function of their minimum duration. Black and red lines correspond to simulations of the MRS and the MRS including revisits, respectively.}
\label{fig:gap-trackin-separation}
\end{figure}

These gaps can be used to plan additional tasks not considered as constraining in the scheduler. This is for example the case of flat field and dark current calibration images, which can be easily fitted in gaps lasting around $\sim$1\,hour. Left panel in Figure\,\ref{fig:gap-trackin-separation} shows the median separation between gaps as a function of their duration for the simulations of the MRS both excluding and including revisits to targets as black and red lines, respectively. In both cases, there are gaps of around 1\,hour every $\sim$12\,hours. This will allow to plan these calibrations very frequently if needed. Alternatively, these short gaps can help to extend the out-of-transit baseline of the observations, which may provide useful information to correct for stellar variability and improve the determination of the transit or occultation depth.

On the other hand, data downlink periods have not been considered in the mission planning because they can be simultaneous to observations. They are expected to take about 14\,hours per week distributed in three periods of 4 or 6\,hours. Left panel in Figure\,\ref{fig:gap-trackin-separation} reveals that these cadence is fulfilled by gaps longer than 4 hours only if targets are not revisited. Otherwise, long gaps are separated by several days. However, in this case, since data downlink can be done simultaneously to exoplanet observations, they can be scheduled when tracking targets longer than 4-6\,hours. Right panel in Figure\,\ref{fig:gap-trackin-separation} shows that such long target pointings occur almost every 0.6\,days.

Finally, gaps longer than 1\,hour in which no additional transit or occultations can be fitted, can also be used for non-time constrained ancillary science such as the observation of brown dwarfs \cite{Morales2015a} or variable young stellar objects. As an example, scheduling simulations show that 90\% of these gaps can be used to follow-up various types of young stellar objects, including FU\,Orionis and T\,Tauri type stars, and systems harbouring extreme debris disks \cite{Gyurus2020}. Targets of opportunity that may rise during \emph{Ariel} nominal operations phase could also be planned within this gaps, if possible, or included in the scheduler to recompute optimized plans.

\section{Conclusions}
\label{sec:conclusions}

The main conclusion of the different simulations of the \emph{Ariel} mission planning is that almost all the targets in the core sample can be observed as requested while meeting all mission and system requirements and constraints. Only targets that are observable few times are challenging, and the scheduler can be used to easily identify them.

Despite the fact that slots of inactive time are inevitable in the \emph{Ariel} context because observations are time-constrained, $\sim$85--90\% of the mission time can be devoted to science, including revisits to targets in the reference sample, observation of back-up targets, or using gaps for ancillary science. In fact, by choosing wisely the targets to re-observe, it is possible to increase the number of well-characterized targets in the tier 3 subsample by a factor of $\sim$2.5, or surveying several more Neptune and Earth-like planets considered as back-up targets. In summary, about 24000 hours can be scheduled on observations of some 3600 transit or eclipse events. In terms of the time that can be used for exoplanet observations, similar conclusions are reached using our previous MOEA \cite{GarciaPiquer2015} or heuristic \cite{Morales2015b} algorithms when revisits to targets are considered in order to maximize the science time.
The difference of the scheduling tool approach presented here is that it produces plans with a better optimization of the number of completed targets, thus improving also the time used for science even if revisits are not considered, and faster than using MOEA algorithms.

The observation of phase curves have a significant impact on the total number of surveyed stars. The results show that 43 of such observations, representing $\sim$11\% of the \emph{Ariel} available time, can be scheduled. This can be done at the expense of reducing the number of tier 1 targets by 69 exoplanets. However, such kind of observations does include the follow-up of transit and occultation and provide a wealth of information about exo-atmospheres. Besides, the test cases presented here demonstrate that a balance between the MRS and phase curve subsample can be found to fulfil the \emph{Ariel} scientific goals. On the other hand, such observations naturally reduces the fraction of waiting because observations are longer, improving the \emph{Ariel} mission efficiency. 

To summarize conclusions, the scheduling algorithm presented here will be an efficient tool to plan \emph{Ariel} observations, just taking few minutes to schedule the whole mission considering all constraints. The scheduling simulations also reveal that the primary mission science goals can be fulfilled within the 3.5 years operational phase, even including the observation of several phase curves. Besides, they also reveal the scheduling tools as a helpful way to select the mission reference sample of exoplanets that best fits the mission plan. The scheduling tool presented here is being continually upgraded to improve the mission plan optimization and to fasten its execution.

\begin{acknowledgements}
This publication has been made possible by grants ESP2016-80435-C2-1-R and PGC2018-098153-B-C33 funded by MCIN/AEI/10.13039/501100011033 and by “ERDF A way of making Europe”. We acknowledge as well the support of the Generalitat de Catalunya/CERCA programme.
\end{acknowledgements}




\end{document}